\documentclass[fleqn,10pt]{wlscirep}

\title{Noise of a superconducting magnetic flux sensor based on a proximity Josephson junction}

\author[1,*]{R. N. Jabdaraghi}
\author[1]{D. S. Golubev}
\author[1]{J. P. Pekola}
\author[1]{J. T. Peltonen}
\affil[1]{Low Temperature Laboratory, Department of Applied Physics, Aalto University School of Science,
P.O. Box 13500, FI-00076 Aalto, Finland}

\affil[*]{robab.najafi.jabdaraghi@aalto.fi}

\newcommand{\mr}[1]{\mathrm{#1}}
\newcommand{\be}{\begin{equation}}
\newcommand{\ee}{\end{equation}}

\newcommand{\onlinecite}[1]{\hspace{-1 ex} \nocite{#1}\citenum{#1}} 

\newcommand{\kb}{k_{\mr{B}}}

\newcommand{\figta}{$\left(\mathrm{a}\right)\;$}
\newcommand{\figtb}{$\left(\mathrm{b}\right)\;$}
\newcommand{\figtc}{$\left(\mathrm{c}\right)\;$}
\newcommand{\figtd}{$\left(\mathrm{d}\right)\;$}
\newcommand{\figte}{$\left(\mathrm{e}\right)\;$}
\newcommand{\figtf}{$\left(\mathrm{f}\right)\;$}

\newcommand{\figa}{$\left(\mathrm{a}\right)$}
\newcommand{\figb}{$\left(\mathrm{b}\right)$}
\newcommand{\figc}{$\left(\mathrm{c}\right)$}

\newcommand{\fige}{$\left(\mathrm{e}\right)$}

\newcommand{\figg}{$\left(\mathrm{g}\right)$}

\newcommand{\mk}{\;\mr{mK}}
\newcommand{\nm}{\;\mr{nm}}
\newcommand{\kelvin}{\;\mr{K}}
\newcommand{\second}{\;\mr{s}}
\newcommand{\na}{\;\mr{nA}}
\newcommand{\pa}{\;\mr{pA}}
\newcommand{\fa}{\;\mr{fA}}
\newcommand{\pf}{\;\mr{pF}}
\newcommand{\nf}{\;\mr{nF}}

\newcommand{\mv}{\;\mr{mV}}

\newcommand{\volt}{\;\mr{V}}
\newcommand{\ohm}{\;\Omega}
\newcommand{\kohm}{\;\mr{k}\Omega}
\newcommand{\muv}{\;\mu\mr{V}}
\newcommand{\muev}{\;\mu e\mr{V}}

\newcommand{\muh}{\;\mu\mr{H}}

\newcommand{\mhz}{\;\mr{MHz}}

\newcommand{\hz}{\;\mr{Hz}}

\newcommand{\nphisqrthz}{\;\mr{n}\Phi_0/\mr{Hz}^{1/2}}
\newcommand{\muphisqrthz}{\;\mu\Phi_0/\mr{Hz}^{1/2}}
\newcommand{\asqhz}{\;\mr{A}^2/\mr{Hz}}

\newcommand{\phio}{\;\Phi_0}
\newcommand{\phii}{\Phi}
\newcommand{\ccoax}{C_{\mr{coax}}}
\newcommand{\ccoup}{C_{\mr{c}}}
\newcommand{\nef}{\mr{NEF}}
\newcommand{\te}{T_{\mr{e}}}
\newcommand{\vsd}{V_{\mr{SD}}}
\newcommand{\isd}{I_{\mr{SD}}}
\newcommand{\rl}{R_{\mr{l}}}
\newcommand{\lline}{L_{\mr{l}}}
\newcommand{\cl}{C_{\mr{l}}}
\newcommand{\rs}{R_{\mr{S}}}
\newcommand{\rt}{R_{\mr{T}}}
\newcommand{\ns}{n_{\mr{S}}}
\newcommand{\nn}{n_{\mr{N}}}
\newcommand{\si}{S_{\mr{I}}}
\newcommand{\sis}{S_{\mr{I}_{\mr{S}}}}
\newcommand{\sir}{S_{\mr{I}_{\mr{R}}}}
\newcommand{\sia}{S_{\mr{I}_{\mr{A}}}}
\newcommand{\sishot}{S_{\mr{I}_{\mr{shot}}}}
\newcommand{\sphi}{S_{\Phi}}
\newcommand{\sva}{S_{\mr{V}_{\mr{A}}}}
\newcommand{\sv}{S_{\mr{V}}}
\newcommand{\flc}{f_0}

\newcommand{\reff}{R_{\mr{eff}}}
\newcommand{\zeff}{Z_{\mr{eff}}}
\newcommand{\zr}{Z_{\mr{R}}}
\newcommand{\ph}{P_{0}}

\newcommand{\vin}{V_{\mr{in}}}
\newcommand{\dis}{\delta I_{\mr{S}}}
\newcommand{\dir}{\delta I_{\mr{R}}}
\newcommand{\dia}{\delta I_{\mr{A}}}
\newcommand{\dva}{\delta V_{\mr{A}}}
\newcommand{\dishot}{\delta I_{\mr{shot}}}

\newcommand{\iphi}{I({{\Phi}})}
\newcommand{\vphi}{V({{\Phi}})}
\newcommand{\dphi}{\delta{{\Phi}}}
\newcommand{\didphimax}{|\partial{I}/\partial{{\Phi}}|_{\mr{max}}}
\newcommand{\dvdphimax}{|\partial{V}/\partial{{\Phi}}|_{\mr{max}}}
\newcommand{\didphi}{\partial{I}/\partial{ \Phi}}
\newcommand{\dvdphi}{\partial{V}/\partial{ \Phi}}

\begin{abstract}
We demonstrate simultaneous measurements of DC transport properties and flux noise of a hybrid superconducting magnetometer based on the proximity effect (superconducting quantum interference proximity transistor, SQUIPT). The noise is probed by a cryogenic amplifier operating in the frequency range of a few MHz. In our non-optimized device, we achieve minimum flux noise $\sim 4\muphisqrthz$, set by the shot noise of the probe tunnel junction. The flux noise performance can be improved by further optimization of the SQUIPT parameters, primarily minimization of the proximity junction length and cross section. Furthermore, the experiment demonstrates that the setup can be used to investigate shot noise in other nonlinear devices with high impedance. This technique opens the opportunity to measure sensitive magnetometers including SQUIPT devices with very low dissipation.
\end{abstract}

\begin{document}
\thispagestyle{empty}
\flushbottom
\maketitle

\section*{ Introduction}

Measuring noise provides an uncompromising test of microscopic and nanoscopic superconducting sensors~\cite{Martinez2016,Schmelz2015,Martinez2017,Schmelz2017,Vasuykov2013}, such as superconducting quantum interference devices (SQUIDs), for ultra-sensitive detection of weak and local magnetic signals. A hybrid superconducting magnetometer~\cite{Petrashov1994,Petrashov1995} based on the proximity effect~\cite{Tinkham1996} (superconducting quantum interference proximity transistor, SQUIPT~\cite{Giazotto2010}) has demonstrated in experiments high responsivity to magnetic flux~\cite{Giazotto2010,Meschke2011,Najafi2014,Ronzani2014} and theoretically~\cite{Giazotto2011} the noise is predicted to be very low, comparable to or below $50\nphisqrthz$ obtained with state-of-the-art nanoSQUIDs~\cite{Schmelz2017,Vasuykov2013}. Yet the intrinsic limits to flux noise performance of such a device have not been experimentally investigated in detail up to now. Here, we present a measurement of flux noise of a SQUIPT using a cryogenic amplifier~\cite{DiCarlo2006,Hashisaka2008,Hashisaka2008Co,Arakawa2013} operating in the frequency range of a few MHz.

A SQUIPT interferometer consists of a superconducting loop interrupted by a short normal-metal wire in direct metal-to-metal contact while an additional superconducting probe electrode is tunnel-coupled to the normal region, cf. Fig.~\ref{fig:setup}~\figa. Its operation relies on the phase dependence of the density of states (DoS) in the normal part~\cite{Belzig1999}, probed via the tunnel junction. The figure of merit of a SQUIPT magnetometer is the noise-equivalent flux ($\nef$) or flux sensitivity~\cite{Likharev1986}, which has been considered theoretically in Ref.~\onlinecite{Giazotto2011}. In the earliest experimental realization~\cite{Giazotto2010}, the $\nef$ was limited by the preamplifier contribution to the noise, and estimated to be $\sim 20\muphisqrthz$. In a subsequent optimized device with a shorter proximity junction, $500\nphisqrthz$ has been obtained at $240\mk$ in a low-frequency (sub-kHz) cross-correlation measurement, still limited by the room-temperature amplifier noise~\cite{Ronzani2014}. Recently, $260\nphisqrthz$ at $1\kelvin$ was reported for a fully superconducting device~\cite{Ronzani2016}. However, the challenging task has remained to observe directly the non-bandwidth-limited intrinsic flux noise performance of the hybrid superconducting magnetometer devices, predicted to be determined by shot noise in the current through the probe tunnel junction~\cite{Giazotto2011}.

Besides hindering sensor operation, the shot noise~\cite{Schottky1918,Blanter2000} in the electrical current of a biased conductor provides information on quantum transport in mesoscopic structures beyond the average current~\cite{Landauer1998}. It has been measured in various systems, including quantum point contacts (QPCs)~\cite{Reznikov1995,Kumar1996} and quantum dots (QDs)~\cite{Zarchin2008}, and found to provide an accurate means of primary thermometry for metallic tunnel junctions~\cite{Spietz2003,Spietz2006} and recently for QPCs as well~\cite{Iftikhar2016}. A successful technique for measuring the shot noise of high-impedance semiconducting samples relies on a cryogenic amplifier based on a high electron mobility transistor (HEMT) and an RLC tank circuit with resonance frequency of a few MHz~\cite{DiCarlo2006,Hashisaka2008,Hashisaka2008Co}. Such an approach avoids the ubiquitous amplifier 1/f noise and signal loss due to low pass filter formed by cable capacitance and the high sample inductance. For enhanced sensitivity, the method extends straightforwardly to cross-correlation of signals from two amplifiers~\cite{DiCarlo2006}, and its adaptations have been employed to study the noise of QPCs~\cite{Nishihara2012,Nakamura2009} and QDs~\cite{Yamauchi2011,Okazaki2013}, including demonstration of the quantum of thermal conductance for heat flow in a single electronic channel~\cite{Jezouin2013}. In this work we use the technique to characterize the flux noise of a hybrid superconducting tunnel junction magnetometer. Here we present simultaneous measurements of the DC transport properties and current noise of a SQUIPT interferometer, and use them to infer a $\nef\approx 4\muphisqrthz$ in the non-optimized structure with a proximity SNS junction of length $l\approx 245\nm$. We show that the low-temperature readout of RLC filtered shot noise can be applied to the study of nonlinear devices once changes in the differential resistance are taken into account, cf. gate-tunable semiconducting devices where the resistance depends only weakly on the bias voltage.

\section*{Results}

\subsection*{Noise measurement setup}
\begin{figure}[tb]
\centering
\includegraphics[width=0.7\columnwidth]{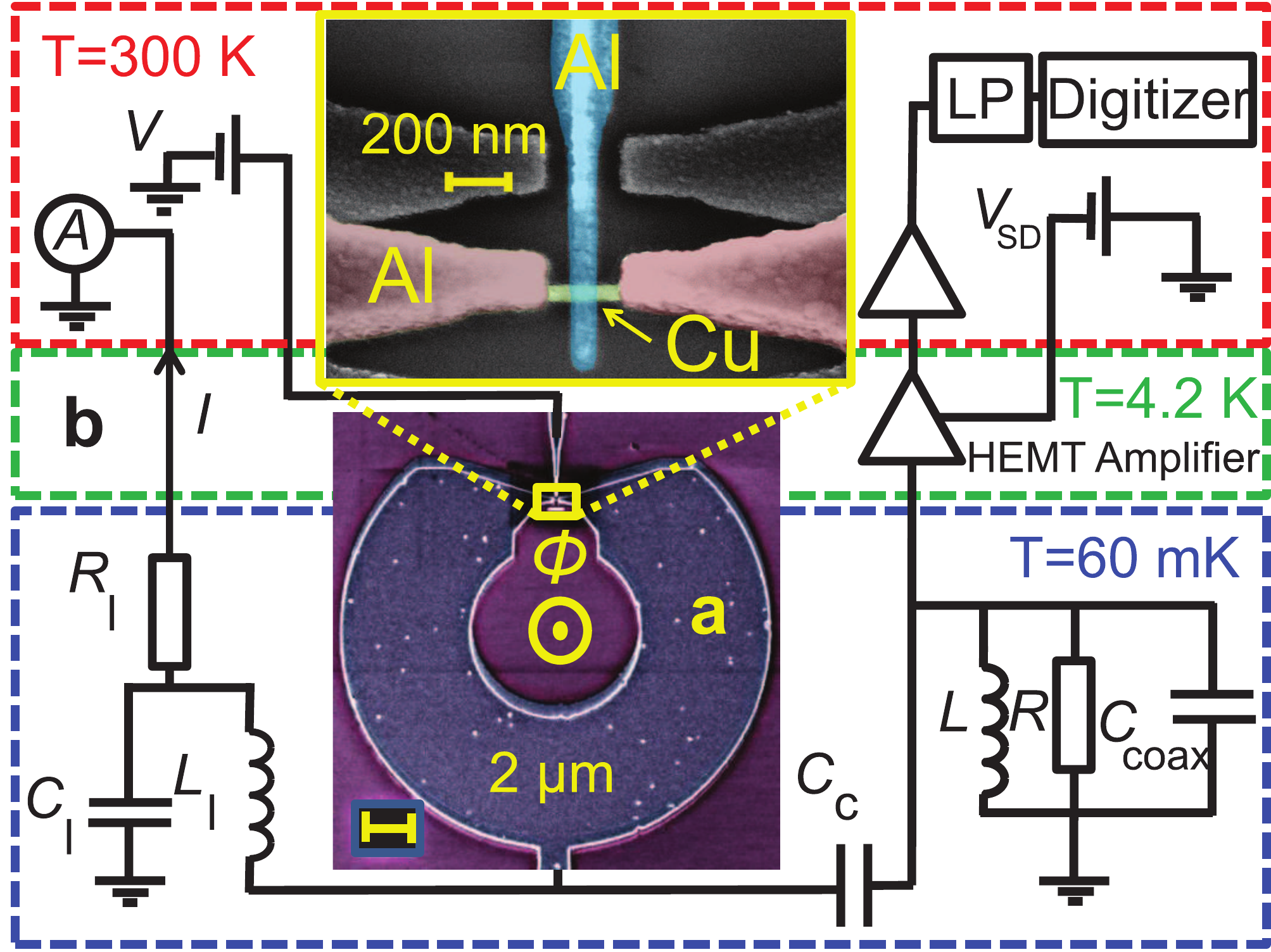}
\caption{Typical sample and the noise measurement setup. \textbf{(a)} False color scanning electron micrograph of the SQUIPT device, together with a zoomed-in view of the Cu island (green) embedded in the superconducting Al loop (brown). The Al tunnel probe (blue) contacts the middle of the proximity SNS junction. \textbf{(b)} Schematic view of the DC and noise measurement system in the dilution refrigerator. Fluctuations of current through the sample are converted to voltage noise at the HEMT amplifier input by a resonant circuit formed mainly by the inductors $L$ and $\lline$ on the sample holder, and the distributed cable capacitance $\ccoax$ that connects the sample to the cryogenic amplifier residing directly in the helium bath.} \label{fig:setup}
\end{figure}

\begin{figure}[tb]
\centering
\includegraphics[width=0.7\columnwidth]{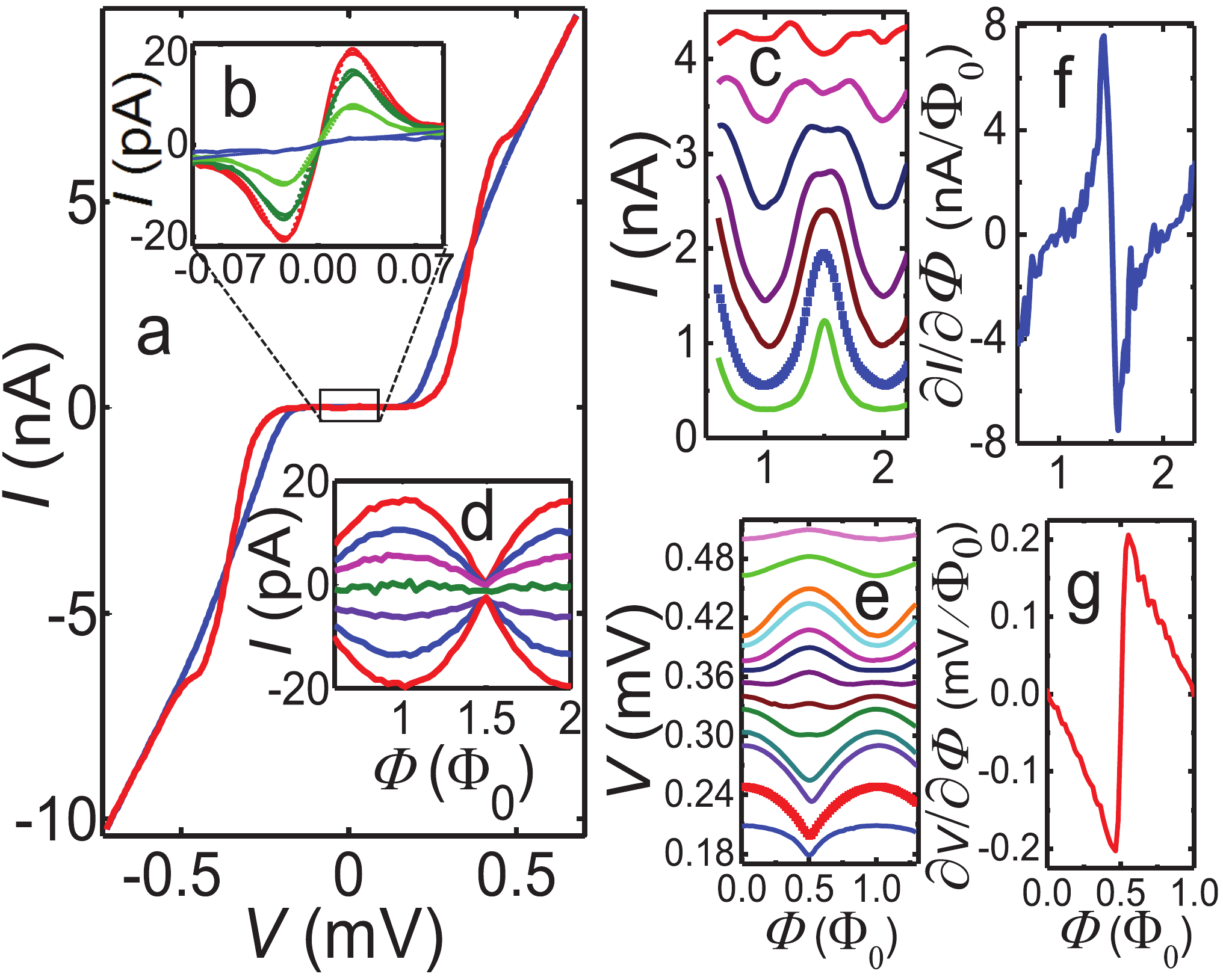}
\caption{DC transport measurements. \textbf{(a)} IV characteristics at $T=60\mk$, measured at two values of magnetic flux $\phii=0$ (red solid line) and $\phii=0.5\phio$ (blue solid line), respectively. \textbf{(b)} Flux modulation of the IV curve (solid lines) around zero bias voltage at several values of magnetic flux between $\phii=0$ and $\phii=0.5\phio$, together with the theoretical model at each flux (dotted lines). \textbf{(c)} Current modulation $\iphi$ at various fixed bias voltages $V\gtrsim\Delta/e$, and \textbf {(d)}, in the sub-gap region close to zero bias voltage. \textbf{(e)} Measured flux-to-voltage curves $\vphi$ at several values of the bias current through the device. \textbf{(f)} Current responsivity $\didphi$ and \textbf{(g)}, voltage responsivity $\dvdphi$ as functions of the magnetic flux at the optimum bias points, $V=0.249\mv$ and $I=4.2\na$, respectively.} \label{fig:IV}
\end{figure}

\begin{figure*}[htb]
\centering
\includegraphics[width=0.7\columnwidth]{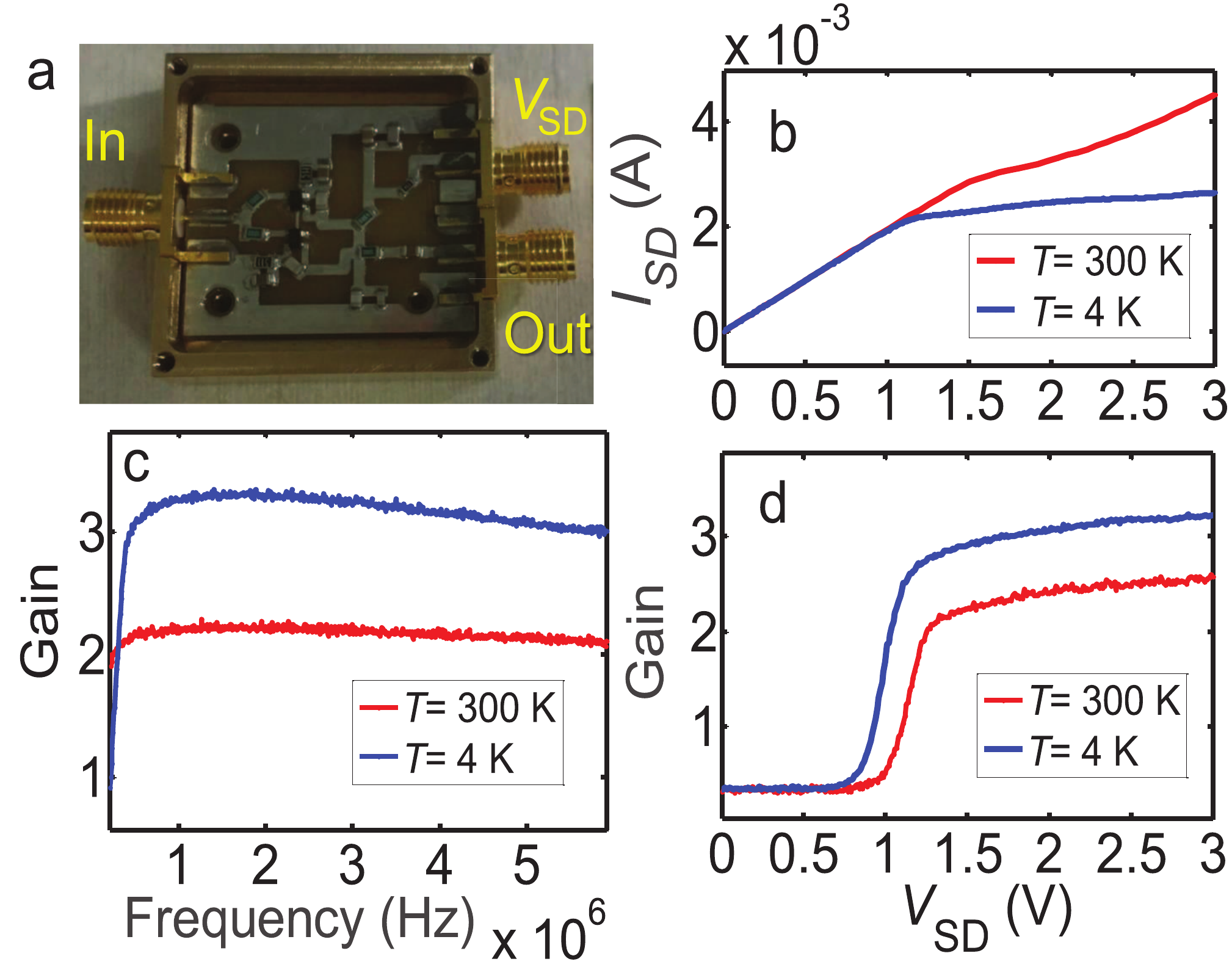}
\caption{Cryogenic HEMT amplifier. \textbf{(a)} Inside view of the cryoamplifier with a pair of Avago ATF-34143 HEMT transistors and passive components including surface mount metal-film resistors and laminated ceramic capacitors. The design follows directly the one introduced in Ref.~\onlinecite{Arakawa2013}. \textbf{(b)} Source-drain current $\isd$ and \textbf{(d)} gain as a function of the supply voltage $\vsd$ at two different temperatures $T=4.2\kelvin$ (blue solid line) and $T=300\kelvin$ (red solid line) at $3\mhz$. By decreasing the temperature, $\isd$ decreases while the gain increases. \textbf{(c)} Frequency dependence of the gain at $\vsd=2\volt$ at two different temperatures.}
\label{fig:HEMT}
\end{figure*}

A typical SQUIPT based on a superconducting aluminium loop placed into a perpendicular magnetic field is presented in Fig.~\ref{fig:setup}~\figa, with an enlarged view of the weak link region depicted in the top inset. It is fabricated using conventional methods of electron beam lithography and metal deposition through a suspended mask (see further fabrication details in the Methods Section). The noise measurement setup is installed in a $^3$He/$^4$He dilution refrigerator with base temperature close to $60\mk$ as shown in Fig.~\ref{fig:setup}~\figb. As the main elements, our home made double-HEMT cryogenic amplifier~\cite{Arakawa2013} (see Fig.~\ref{fig:HEMT} in the Methods Section for amplifier characterization at room temperature and $4.2\kelvin$) and the inductors of the LC resonant circuit are placed in the liquid helium bath and on the sample holder at base temperature, respectively. The voltage source $\vsd$ is used to bias the amplifier. A bias voltage $V$ is applied to the SQUIPT tunnel probe electrode, and the average current $I$ is measured with a room-temperature current amplifier through the line with inductor $\lline$. This line is low-pass filtered by the resistance $\rl=330\ohm$ and capacitance $\cl=22\nf$. An identical filter is included in the biasing line of the tunnel probe but omitted in Fig.~\ref{fig:setup}~\figtb for clarity.

Simultaneously with measurement of the average current $I$, current noise through the SQUIPT is probed by the HEMT amplifier via the capacitor $\ccoup$. At frequencies of the order of the resonance at $\flc=1/(2\pi\sqrt{L'\ccoax})\approx 4.2\mhz$, formed by the inductance $L'=(L^{-1}+\lline^{-1})^{-1}\approx 16.5\muh$ (due to the coils $L=\lline=33\muh$ on the sample holder) and the capacitance $\ccoax\approx 92\pf$ (mainly due to distributed cable capacitance between the sample holder and the amplifier), the capacitors $\ccoup=\cl$ can be considered as electrical shorts. Importantly, this results in a robust peak signature of the white shot noise of the sample, filtered by the characteristic band-pass response of the $RL'\ccoax$ circuit, to be present in the observed voltage noise spectral density. In Fig.~\ref{fig:setup}~\figtb, the phenomenological resistor $R\gtrsim 50\kohm$ denotes the parasitic losses in the circuit, mainly the  inductors $L$ and $\lline$. It accounts for the losses in the circuit when the differential resistance of the sample $\rs(V,\phii)=dV/dI\gtrsim R$. The signal is further amplified by another stage (SRS SR445A) at room temperature, and low pass (LP) filtered by a commercial $5\mhz$ filter to avoid aliasing. The amplified voltage signal is finally captured by a 16-bit digitizer running continuously at 50 MSamples/s, converted into spectral density of voltage noise by windowing and Fast Fourier Transform of blocks with typically $2^{15}$ samples~\cite{DiCarlo2006}, and a desired number of spectra are averaged together to improve the signal-to-noise ratio.

\begin{figure*}[tb]
\centering
\includegraphics[width=\columnwidth]{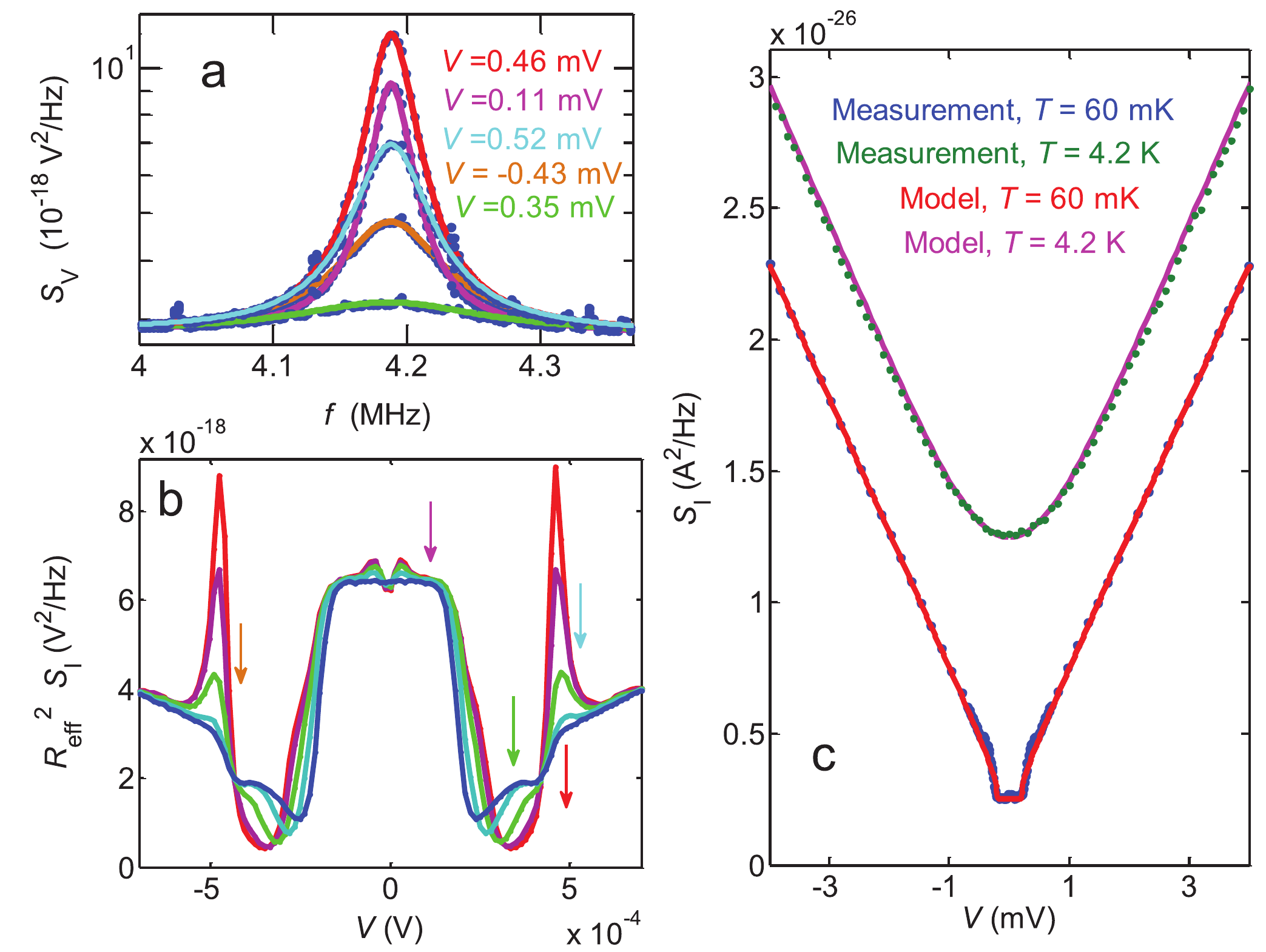}
\caption{Noise measurements. \textbf{(a)} Power spectral density of the measured voltage noise at ${{\Phi}}=0$ for the indicated values of the bias voltage $V$ (blue dots). The solid lines are fits to Eq.~\ref{sv}. \textbf {(b)} Bias dependence of the peak height $\reff^2\si$, extracted from fits to Eq.~\ref{sv}, for a few equally spaced values of magnetic flux between $\phii=0$ (red solid line) and $\phii=0.5\phio$ (blue solid line). \textbf {(c)} Total current noise $\si$ vs. the DC bias, measured at $T=4.2\kelvin$ with the junction in the normal state, and at $T=60\mk$ in the superconducting state, together with the theoretical predictions (see text for details).} \label{fig:noisefit}
\end{figure*}

\subsection*{DC transport measurements}

Figure~\ref{fig:IV}~\figta displays the experimental current--voltage (IV) characteristics of the device recorded at $T=60\mk$ at two different magnetic fields, $\phii=0$ and $\phii=0.5\phio$, which correspond to maximum and minimum minigap opened in the normal metal DoS~\cite{Sueur2008,Zhou1998}, respectively. At large biases $|V|\gtrsim 0.5\mv$ the resistance of the tunnel junction approaches the asymptotic normal-state value $\rt\approx 60\kohm$. Figure~\ref{fig:IV}~\figtb further shows an enlarged view of the flux dependence of the sub-gap current. Full phase modulation, i.e., complete suppression of the supercurrent at $\phii=0.5\phio$, is observed due to the small Al loop inductance compared to that of the SNS weak link~\cite{Najafi2014,Ronzani2014}. The shape of the supercurrent peaks shows good agreement with a theoretical calculation (dotted lines) based on the $P(E)$ theory of incoherent Cooper pair tunneling~\cite{Averin1990,Ingold1992}, assuming the junction to be embedded in an effective $RC$ environment.

We next characterize the flux responsivity of the SQUIPT device by measuring current $\iphi$ and voltage $\vphi$ modulations at different values of bias voltage or current applied to the tunnel probe. Figures~\ref{fig:IV}~\figtc and~\ref{fig:IV}~\figte illustrate some of such current and voltage modulations in the bias range from $0.246\mv$ to $0.369\mv$ and $0.14\na$ to $7\na$, respectively. Furthermore, Fig.~\ref{fig:IV}~\figtd shows the measured current modulation at several sub-gap bias voltages. With the $\iphi$ and $\vphi$ characteristics at hand, we obtain the flux-to-voltage transfer function $\dvdphi$ and flux-to-current transfer function $\didphi$ by numerical differentiation. The maximum absolute values $\didphimax\simeq 8\na/\Phi_0$ and $\dvdphimax\simeq 0.2\mv/\Phi_0$ are reached at $V\approx 249\muv$ and $I\approx 4.2\na$, respectively. The transfer functions close to these optimum bias values are plotted in Figs.~\ref{fig:IV}~\figtf and~\ref{fig:IV}~\figg, whereas the corresponding $\iphi$ and $\vphi$ characteristics are shown in bold in Figs.~\ref{fig:IV}~\figtc and~\ref{fig:IV}~\fige.

\subsection*{Shot noise measurements}

\begin{figure}[tb]
\centering
\includegraphics[width=0.5\columnwidth]{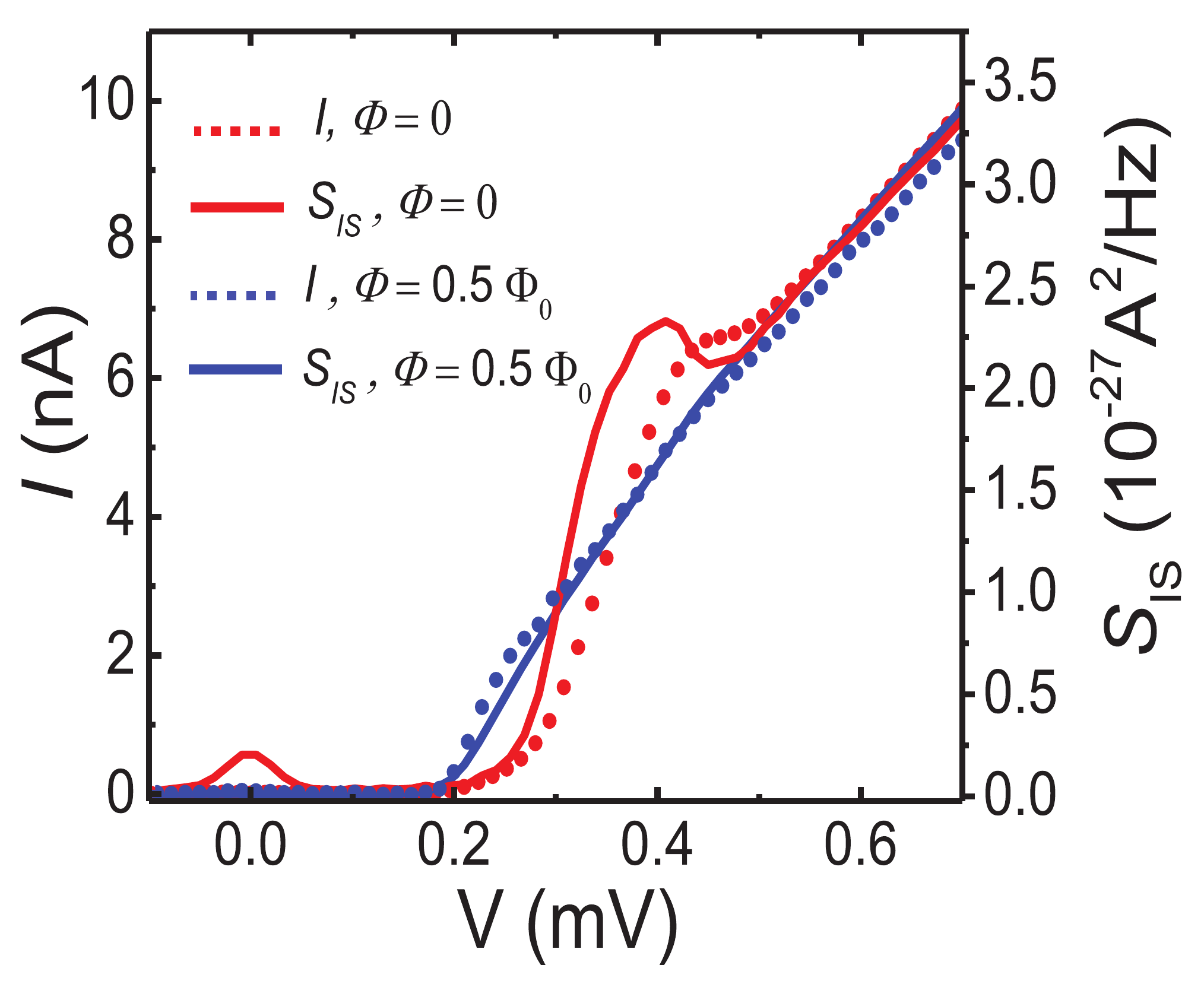}
\caption{Bias dependence of the current noise. IV characteristics of the SQUIPT (red and blue dots) compared to the measured current noise $\sis$ (red and blue solid lines) vs. bias voltage, at the two extreme flux values, $\phii=0$ and $\phii=0.5\phio$.} \label{fig:noisevsbias}
\end{figure}

\begin{figure*}[htb]
\centering
\includegraphics[width=0.75\columnwidth]{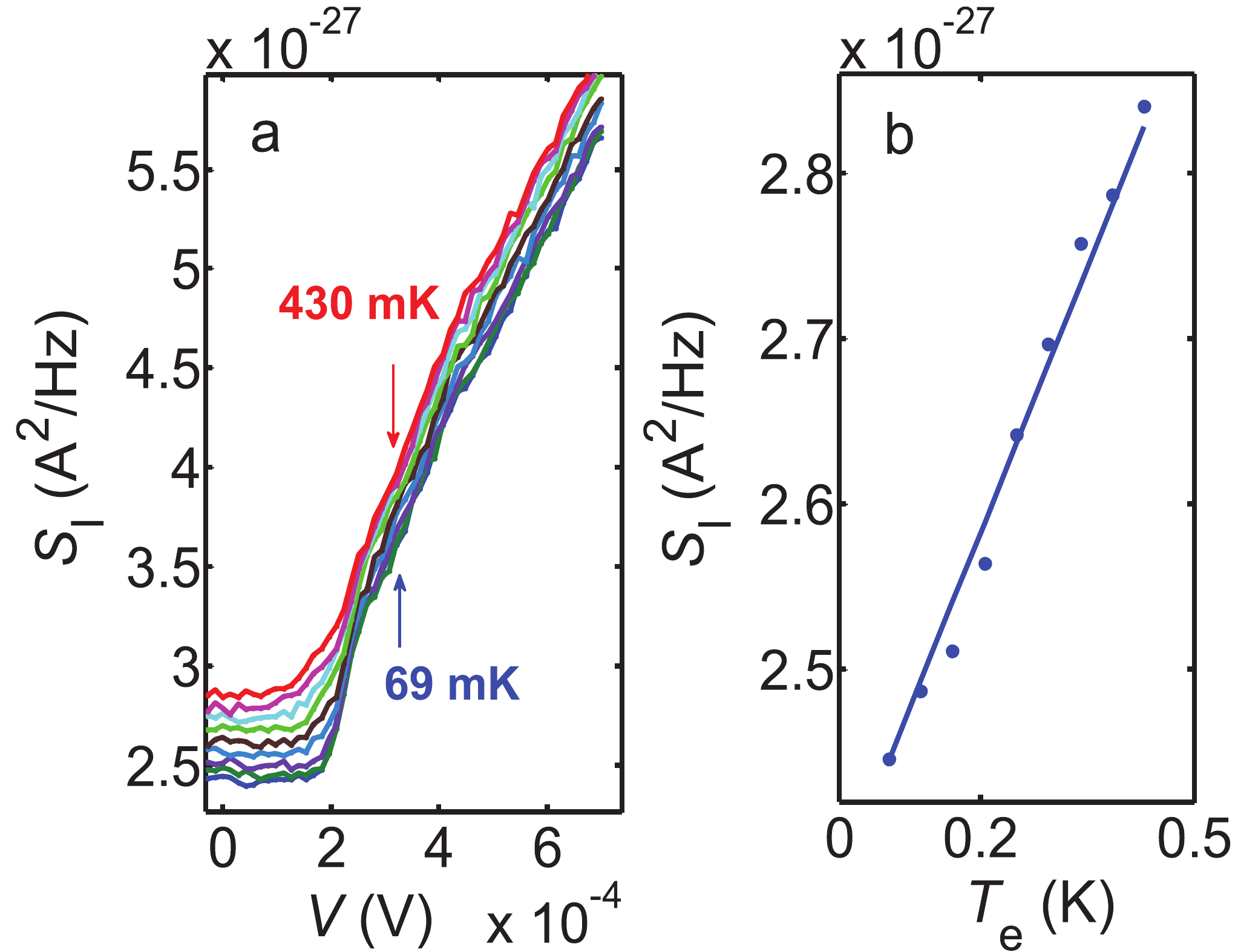}
\caption{Temperature dependence of the total current noise. Besides the measurements at $T=4.2\kelvin$ and $T=60\mk$ discussed in the main text, we have probed $\si$ at $\phii\approx 0.5\phio$ in a range of bath temperatures below $500\mk$. \textbf{a}, Bias dependence of the total noise $\si$ at various values of the bath temperature at $\phii\approx 0.5\phio$. \textbf{b}, Temperature dependence of $\si$ at $V=0$ and $\phii\approx 0.5\phio$. As expected for thermal noise of $\reff$ ($\approx R$ at $V=0$), the signal increases approximately linearly with slope $4\kb/R$ (solid blue line) with increasing $T$ on top of a constant background $\sia\approx 2.3\times10^{-27}\asqhz$.} \label{fig:TdT}
\end{figure*}

We now turn to a description of the SQUIPT noise measurements. The blue dots in Fig.~\ref{fig:noisefit}~\figta show examples of measured spectral densities of voltage noise, referred to the HEMT amplifier input. They were recorded at the base temperature with fixed magnetic flux $\phii\approx 0$ through the SQUIPT loop, at the few indicated values of bias voltage $V$ across the device. The solid lines result from nonlinear least squares fitting to~\cite{DiCarlo2006}
\be
\sv(f)=\sva+\frac{\reff^2\si}{1+(f^2-\flc^2)^2/(f\Delta f)^2}, \label{sv}
\ee
showcasing how the white current noise $\si$ is filtered by the bandpass response of the RLC circuit, centered around $\flc$ (refer to Fig.~\ref{fig:circuit} and subsequent discussion in the Methods Section for details). Above, $\sva$ is the input voltage noise of the amplifier, $\reff=(\rs^{-1}+R^{-1})^{-1}$ is the effective resistance in the circuit, and $\si=\sis+\sir+\sia$ denotes the total current noise, composed of the current fluctuations of the sample ($\sis$), equilibrium noise of the parasitic resistance $R$ ($\sir=4\kb T/R$), and a background term ($\sia$), attributed to the amplifier current noise (see also Figs.~\ref{fig:SI_SQUIPT_0}--\ref{fig:TdT} and related discussion in the Methods Section). The current noise of the sample $\sis=\sishot+(\didphi)^2\sphi$ can be further separated into shot noise $\sishot$ in the quasiparticle tunneling current, and external flux noise $\sphi$ mediated by the transfer function $\didphi$. Given the responsivity of the present sample, in our setup with $\flc$ in the MHz regime we expect the second term to be negligible. In the experiment, we have investigated the dependence of $\sv$ and hence $\si$ on $V$, $\phii$, and $T$.

We make the fits to Eq.~\ref{sv} using $\sva$, $\flc$, the peak height $\ph=\reff^2\si$, and the peak width $\Delta f=2\pi L'\flc^2/\reff$ as adjustable parameters. Here, $\Delta f$ gives directly the full width at half maximum (FWHM) of the peak in $\sv$ in the limit $\flc\gg\Delta f$. Of the four parameters, the background level $\sva\approx 3\times 10^{-18}\;\mr{V}^2/\mr{Hz}$ due to the amplifier voltage noise, and the resonance frequency $\flc\approx 4.18\mhz$ can be kept fixed, whereas the peak height and width depend systematically on $V$, $\phii$, and $T$. With the fitting procedure established, Fig.~\ref{fig:noisefit}~\figtb demonstrates typical bias dependence of the extracted values of the peak height $\ph$. The different curves correspond to a few equally spaced flux values between $\phii=0$ and $\phii=0.5\phio$, whereas the vertical arrows indicate the bias voltages at $\phii=0$ for the spectra displayed in panel~\figa. It is notable that both the peak width $\Delta f$ and height $\ph$ reflect strongly the bias- and flux-dependent changes in $\reff$ and hence $\rs(V,\phii)$ while the noise $\si\propto\ph\Delta f^2$ calculated from these parameters follows $\si\propto|\iphi|$, highlighting the contribution of the shot noise in the tunnel junction.

Figure~\ref{fig:noisefit}~\figtc shows the bias dependence of the total current noise $\si$ extracted in the above manner from noise spectra similar to those in panel~\figa. The two curves correspond to measurements at bath temperature $T=4.2\kelvin$ with the SQUIPT fully in the normal state (top), and at the base temperature $T=60\mk$ (bottom) at a constant magnetic flux. At $T=4.2\kelvin$, the measured noise is well explained by assuming $\sis=(2eV/\rt)\coth(eV/2\kb T)$, shown by the pink solid line, see, e.g., Ref.~\onlinecite{Blanter2000} and references therein. We use the high bias shot noise, i.e., the linear asymptotic increase of $\si$ with $V$, to calibrate the total gain of the setup, by requiring that the slope of $\si$ vs. $V$ equals $2e|I|$. The value is in reasonable agreement with the nominal amplifier gains and expected losses in the circuit. At $T=60\mk$ we also find the noise to be dominated by the shot noise of the SQUIPT tunnel junction: Despite the nonlinear IV at $V\lesssim\Delta/e$, with increasing $V$ the noise increases as $\sis\approx2e|\iphi|$. Here at $T=60\mk$, for the theoretical model for simplicity we use the noise of an NIS tunnel junction, approximately valid for a SQUIPT at magnetic flux $\phii=0.5\phio$ in which case the minigap in the normal metal vanishes~\cite{Sueur2008,Zhou1998} (cf. subsection ``Quasiparticle current fluctuations in a hybrid tunnel junction'' under Methods). As evident in Fig.~\ref{fig:noisefit}~\figc, at base temperature the expected $\sir\approx 0.7\times 10^{-29}\asqhz$ is much smaller than the background term $\sia\approx 2.3\times10^{-27}\asqhz\approx (48\fa)^2/\mr{Hz}$. The origin of the large background current noise requires further study in future work: it is approximately an order of magnitude larger than the amplifier current noise $\sia\approx (13\;\mr{fA})^2/\mr{Hz}$ found in
Ref.~\onlinecite{Arakawa2013}.

In Fig.~\ref{fig:noisevsbias} we show the bias dependence of the noise in more detail at the two extreme flux values $\phii=0$ and $\phii=0.5\phio$, noting that the shot noise directly reflects changes in the average current $I(V,\phii)$. For an explicit comparison with the average current, we plot the corresponding IV characteristics in the same panel, showing that indeed $\sis\approx 2e|I|$. In particular this is well satisfied at $\phii=0.5\phio$. In the $\sis$ curve at $\phii=0$, we attribute the apparent excess noise around zero bias (at the gap edge) to an uncertainty in the fitting to extract the exact value of $\reff$ when the peak is at its narrowest (lowest height). It originates from the residual interfering peaks in the background noise of $\sv$, present for example at $f\approx 4.02\mhz$ in Fig.~\ref{fig:noisefit}~\figa.

\subsection*{Flux noise characterization}

\begin{figure*}[htb]
\centering
\includegraphics[width=0.9\columnwidth]{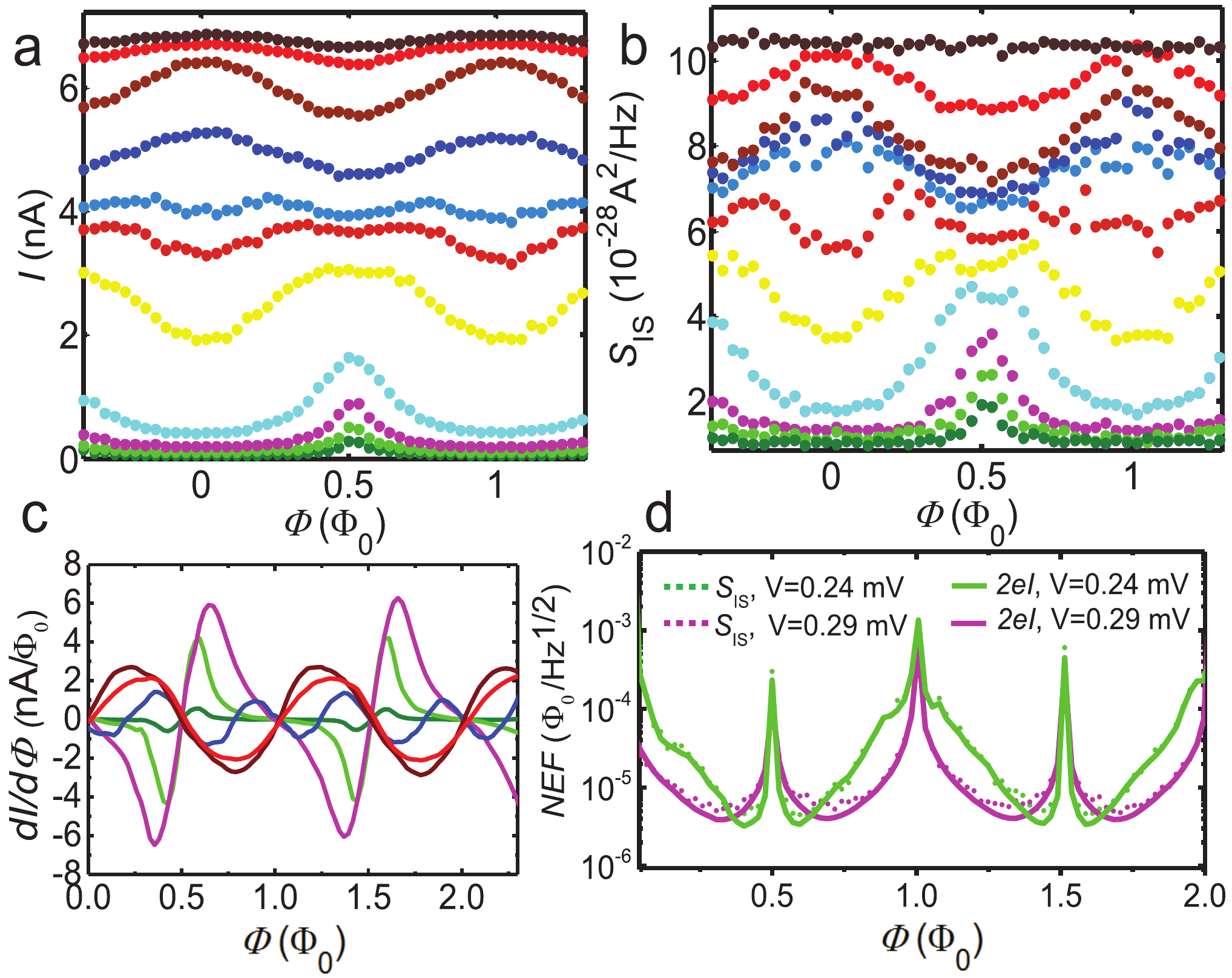}
\caption{Characterization of flux noise performance of the voltage-biased SQUIPT device. \textbf{(a)} DC current and \textbf{(b)}, current noise as a function of magnetic flux at several values of bias voltage, measured simultaneously in the same setup at $60\mk$. \textbf{(c)} Responsivity $\didphi$ at several biases. \textbf{(d)} Flux sensitivity at the 2nd and 3rd lowest bias voltages $V=0.24\mv$ and $0.29\mv$ in panels \figa--\figc. The dotted lines use $\sis$ obtained by direct fitting of the measured $\sv$ spectra, whereas the solid lines assume full shot noise $\sis=2e|\iphi|$ with $\iphi$ from the DC measurement. Each color in panels \figa--\figtd corresponds to a specific bias voltage.} \label{fig:noisevsflux}
\end{figure*}

Two basic figures of merit of the SQUIPT are the transfer function $\didphi$ and the noise-equivalent flux $\nef=\sis^{1/2}/|\didphi|$ (flux sensitivity)~\cite{Likharev1986}. To characterize the flux sensitivity of the device, we measure the flux dependence of $\iphi$ and $\sis(\phii)$ simultaneously in the same setup at several bias voltages $V$ around the onset of the quasiparticle current. Examples of the resulting periodic modulations of $I$ and $\sis$ are shown in Figs.~\ref{fig:noisevsflux}~\figta and~\figb, respectively, demonstrating good qualitative agreement with $\sis\propto|\iphi|$. Figure~\ref{fig:noisevsflux}~\figtc further plots the flux-to-current transfer function $\didphi$, again obtained by numerical differentiation of $\iphi$. With $\sis$ and $\didphi$ at hand, we obtain the $\nef$ curves shown in Fig.~\ref{fig:noisevsflux}~\figtd for two bias values: $V=0.24\mv$ resulting in the lowest $\nef$ (green), and $V=0.29\mv$ giving the highest $\didphi$. The dotted lines use $\sis$ obtained by direct fitting of the measured $\sv$ spectra as discussed above. On the other hand, considering the uncertainties in the fitting procedure, the solid lines assume full shot noise $\sis=2e|\iphi|$ with $\iphi$ taken from the DC measurement. A reasonable fit to the $\sv$ spectra is obtained also under this assumption, resulting only in slight changes in the fitted values of $\reff$.

We achieve the minimum value of $\nef\approx 4\muphisqrthz$ (green solid / dotted line) at the optimum working point $V=0.24\mv$, $\phii\approx 0.4\phio$. For comparison, low frequency ($f\sim 100\hz$) flux noise $\nef\approx 0.5\muphisqrthz$ has been reported for a current-biased, optimized Al-SQUIPT in a room-temperature cross correlation setup~\cite{Ronzani2014}. Likewise, improved flux sensitivity figures down to $\approx 50\nphisqrthz$ have been recently reported for nanoSQUIDs~\cite{Martinez2016,Schmelz2015,Martinez2017,Schmelz2017,Vasuykov2013}. Significant improvements to our initial demonstration of the MHz-range SQUIPT noise performance are expected to result from optimizing the geometry of the device and the consequently enhanced responsivity~\cite{Giazotto2011,Najafi2014,Ronzani2014}. For example, fabricating the interferometer loop from a larger-gap superconductor~\cite{Najafi2016} and making a shorter normal metal wire~\cite{Najafi2014,Ronzani2014}, the transfer function can be enhanced by a few orders up to ${\mu\mr{A}/\Phi_0}$\;~\cite{Ronzani2014} under voltage bias, and flux noise in the $\nphisqrthz$ range has been predicted~\cite{Giazotto2010,Giazotto2011}. Similarly, for devices which reach the maximum responsivity in the supercurrent branch~\cite{Najafi2016}, we expect minimum values of the flux noise in the range of $50\nphisqrthz$. Even higher-bandwidth readout of SQUIPT detectors, similar to fast NIS tunnel junction thermometers~\cite{Schmidt2003,Gasparinetti2015,Saira2016}, is possible by embedding the device in a lumped element or coplanar waveguide resonator with resonance frequency in the range of several hundred MHz or several GHz, respectively.

\section*{Discussion}

In summary, we have investigated the flux noise performance of a SQUIPT interferometer based on shot noise measurements with a cryogenic amplifier at frequencies of the order of a few MHz. This represents the first noise study of such a hybrid interferometer not limited by the low-bandwidth room-temperature readout. The setup is capable of resolving the shot noise of a current $I\sim 100\pa$ in a typical probe junction in an averaging time of the order of $30\second$. In future work, the performance can be further improved by employing a lower-noise room-temperature amplifier, and by using the cross-correlation of signals from two low-temperature amplifiers to reject the uncorrelated background $\sva$ while reliably picking out the signal due to $\sis$. In the present device we reach shot-noise-limited flux sensitivity of the order of $\muphisqrthz$, which can be significantly improved upon optimizing the dimensions of the SNS weak link and the readout tunnel probe.

\section*{Methods}

\subsection*{Fabrication details}
The sample is fabricated using electron beam lithography (EBL) and electron beam evaporation of the Al and Cu thin films. A single lithography step relying on a Ge based hard mask is used to define patterns for multi-angle shadow evaporation of the NIS tunnel probe and the proximity SNS weak link in a single vacuum cycle. The starting point is an oxidized Si substrate onto which we first spin coat a $900\nm$ thick layer of P(MMA-MAA) copolymer. Subsequently, a $22\nm$ thick film of Ge is deposited by electron beam evaporation, followed by spin coating a $50\nm$ thick polymethyl methacrylate (PMMA) layer. The EBL step is followed by first developing the chip in 1:3 solution of methyl isobutyl ketone (MIBK) and isopropanol (IPA) for $30\second$, rinsing in IPA and drying. Reactive ion etching (RIE) with $\mr{CF}_{4}$ (for Ge) and $\mr{O}_{2}$ (for the copolymer layer) is then used to create a suspended mask with proper undercut profile for shadow evaporation. The metals are deposited by electron-beam evaporation: first, $25\nm$ of Al is deposited and oxidized in-situ for 1 min with pure oxygen pressure of 1 millibar to form the tunnel barrier of the normal metal-insulator-superconductor (NIS) probe. Next, approximately $15\nm$ of copper is evaporated to complete the NIS junction and to form the normal metal part of the SNS proximity weak link. Immediately after this, the superconducting Al loop with $120\nm$ thickness is deposited to form clean contacts to the copper island, which completes the structure. Figure~\ref{fig:setup}~\figta shows an SEM image of a resulting SQUIPT device, illustrating the thick Al loop interrupted by the short Cu wire, as well as the thin Al tunnel probe electrode in the middle.

\clearpage

\subsection*{Cryogenic HEMT amplifier}

\clearpage

\subsection*{Model for evaluating the spectrum of voltage noise}

\begin{figure}[htb]
\centering
\includegraphics[width=0.6\columnwidth]{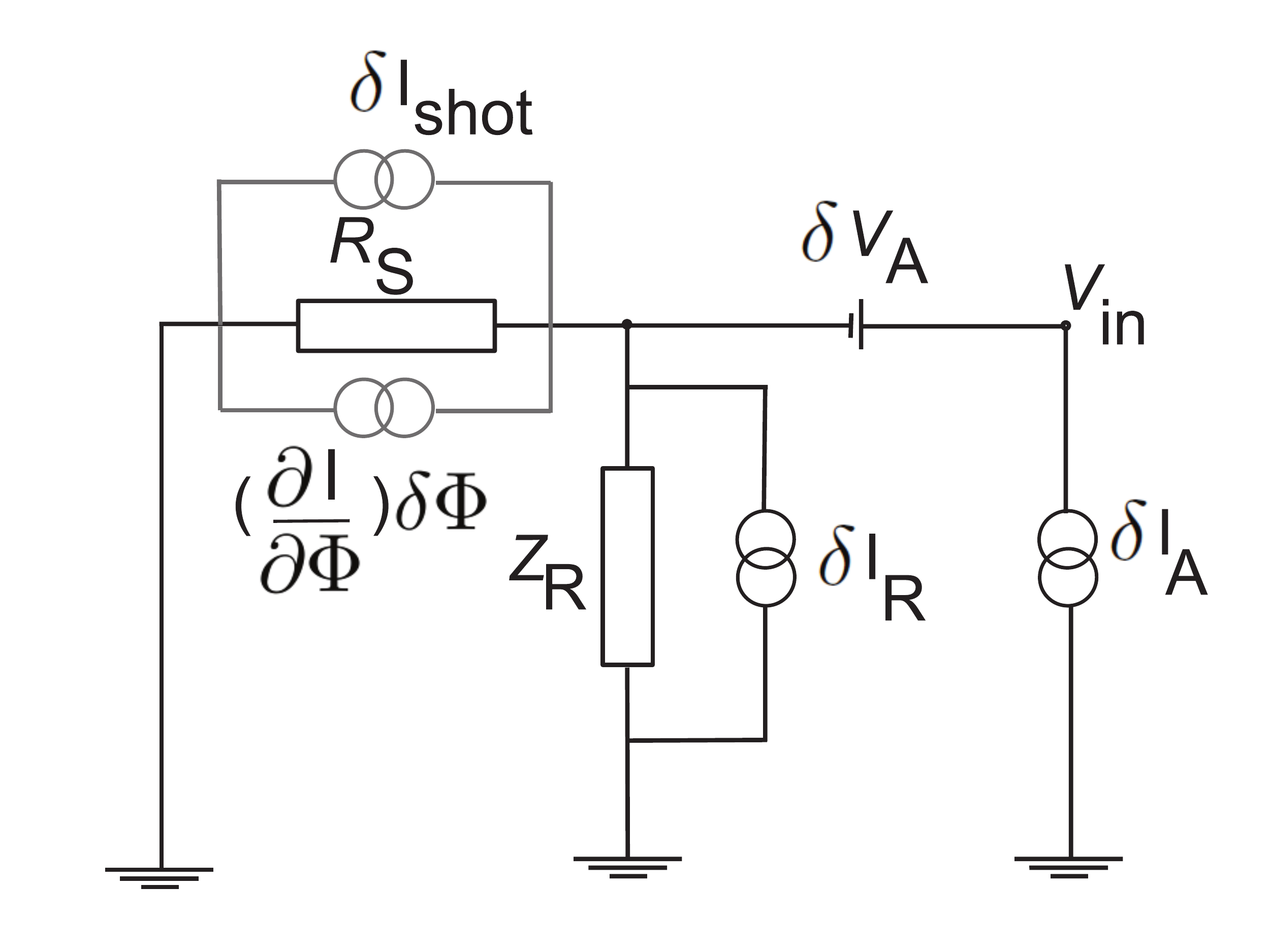}
\caption{Circuit model for evaluating the voltage fluctuations referred to the amplifier input.} \label{fig:circuit}
\end{figure}

\noindent
Figure~\ref{fig:circuit} shows a simplified circuit model of the setup in Fig.~\ref{fig:setup}~\figb, including relevant noise sources for calculating the total voltage noise probed by the HEMT amplifier at its input. Here, it is assumed that the capacitors $\ccoup$ and $\cl$ behave as shorts at frequencies close to $\flc$, whereas $\zr$ denotes the impedance of the parallel RLC circuit, defined via
\be
1/\zr(\omega)=1/R+1/(i\omega L')+i\omega\ccoax, \label{zr}
\ee
with $\omega=2\pi f$. The cryogenic amplifier probes the voltage $\vin$, applied to the gate of its HEMT transistor. The amplifier input voltage and current noise spectral densities are denoted by $\sva$ and $\sia$, respectively. In the following they will be assumed to be white at the frequencies of interest $f\sim\flc$. $\dva$ and $\dia$ represent the corresponding voltage and current noise sources. In Fig.~\ref{fig:circuit}, the input impedance of the amplifier is assumed to be high. The equilibrium current fluctuations in the RLC circuit (i.e., the resistance $R$) are denoted by $\dir$, with spectral density $\sir$. For the total sample noise we write $\dis=\dishot+(\didphi)\dphi$, corresponding to $\sis=\sishot+(\didphi)^2\sphi$.

It is now straightforward to write down Kirchhoff's laws for the circuit. Considering the voltage fluctuation $\Delta\vin(\omega)$ at the amplifier input, we have
\be
\Delta\vin(\omega)=\zeff(\omega)[\dis+\dir+\dia]+\dva. \label{dvin}
\ee
Here $\zeff(\omega)$ is the parallel impedance of the sample and the RLC circuit
\be
1/\zeff(\omega)=1/\rs+1/\zr(\omega). \label{zeff}
\ee
Equation~\ref{dvin} now directly yields the spectral density of the total voltage fluctuations at the amplifier input as
\be
\sv(\omega)=\zeff(\omega)\zeff(-\omega)[\sis+\sir+\sia]+\sva. \label{svin}
\ee
Noting the definition of the effective resistance $\reff=(\rs^{-1}+R^{-1})^{-1}$ and Eq.~\ref{zr}, we see that Eq.~\ref{svin} can be explicitly rewritten to obtain Eq.~\ref{sv} in the main text.

\subsection*{Quasiparticle current fluctuations in a hybrid tunnel junction}

\begin{figure*}[htb]
\centering
\includegraphics[width=0.7\columnwidth]{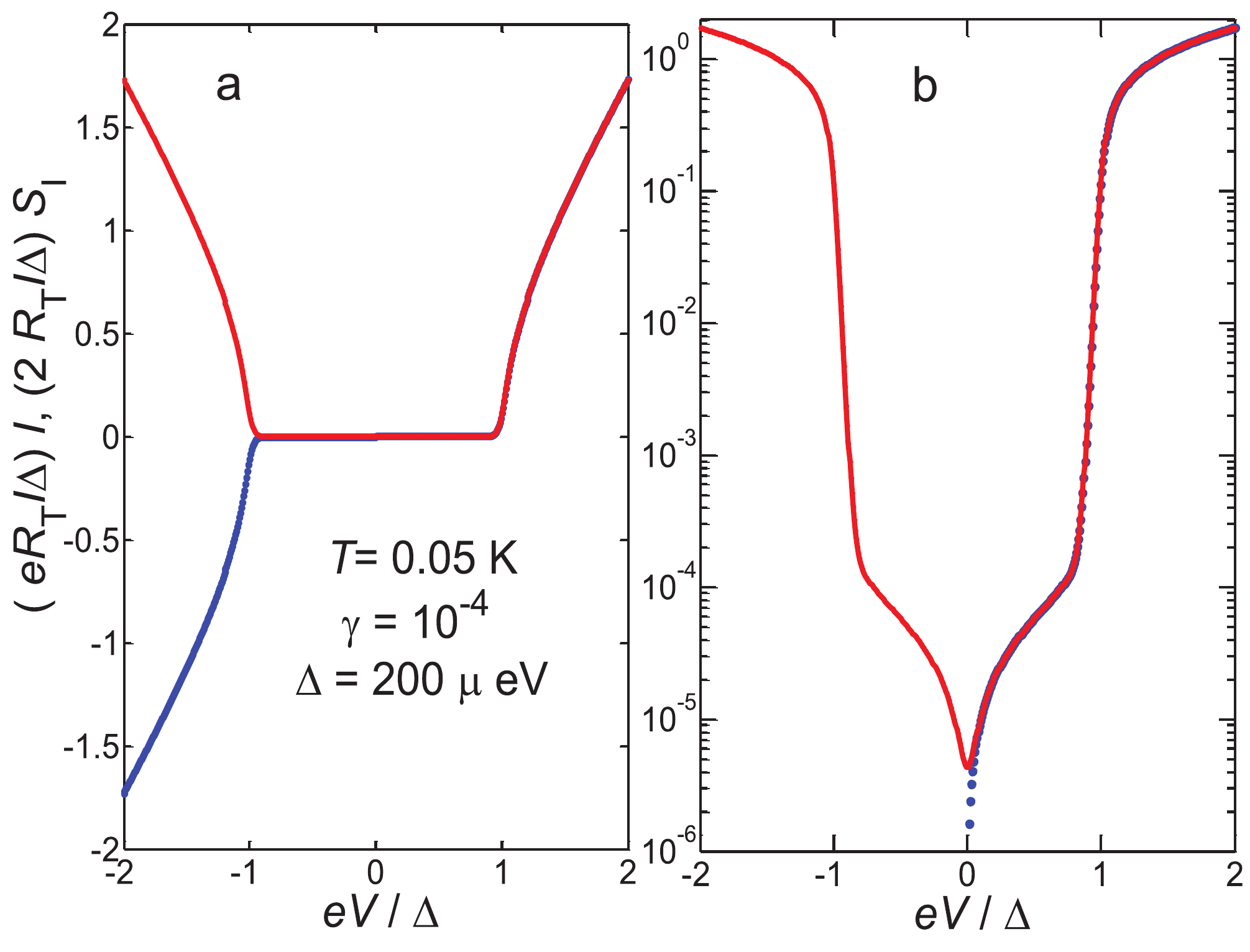}
\caption{Current noise of a NIS junction. \textbf{(a)} Normalized IV characteristics of a NIS tunnel junction (blue dashed line) together with the current noise $\sishot$ (red solid lines) vs. bias voltage calculated at $T=0.05\kelvin$, with the superconducting Al gap $\Delta=200\muev$. Here, the dimensionless parameter $\gamma$ is the ratio between NIS junction asymptotic resistance at high bias voltage and the sub-gap resistance, used in the modeling of a smeared BCS density of states \mbox{$\ns(E)=|\mr{Re}[(E/\Delta+i\gamma)/\sqrt{(E/\Delta+i\gamma)^2-1}]|$}. \textbf{(b)} $I$ and $\sishot$ as in~\figta but plotted on a semilogarithmic scale.} \label{fig:SI_SQUIPT_0}
\end{figure*}

Here we show that despite the non-constant densities of states in both electrodes of the SQUIPT tunnel junction and the nonlinear IV characteristic, the simple approximation $\sishot\approx 2e|I|$ still holds down to relatively low sub-gap bias voltages $V$. In the SNS proximity junction, the density of states $\nn(\epsilon,\phi)$ in the proximized normal metal depends on the phase difference $\phi$ between the S electrodes. This phase- and hence flux-dependent DoS is probed by a tunnel junction with a pure superconducting counterelectrode with the BCS DoS $\ns(\epsilon)$, biased by voltage $V$. Starting from a generic tunnel Hamiltonian, the current noise for a SQUIPT with tunnel resistance $\rt$ can be written as

\be
\centering
\sishot(\omega,V,\phi)=2\rt^{-1}\int d\epsilon \,\nn(\epsilon,\phi)\,\ns(\epsilon-eV)
\{f(\epsilon-eV)[1-f(\epsilon+\hbar\omega)]+[1-f(\epsilon-eV)]f(\epsilon-\hbar\omega)\}. \label{SI_SQUIPT}
\ee
Here we assume a narrow probe electrode and neglect the dependence of $\nn(\epsilon,\phi)$ on the position along the SNS junction~\cite{Meschke2011,Giazotto2011}. In Eq.~\ref{SI_SQUIPT}, $f(\epsilon)=1/[\exp(\epsilon/\kb\te)+1]$ denotes the Fermi-Dirac (quasi-) equilibrium distribution function where $\te$ and $\kb$ are electron temperature and Boltzmann constant, respectively. We further assume the low frequency limit $\hbar\omega \ll\kb \te, eV, \Delta$, yielding
\be
\sishot(V,\phi)=2\rt^{-1}\int d\epsilon \,\nn(\epsilon,\phi)\,\ns(\epsilon-eV) \{f(\epsilon-eV)[1-f(\epsilon)]+[1-f(\epsilon-eV)]f(\epsilon)\}. \label{SI_SQUIPT_0}
\ee
For simplicity, let us consider the SQUIPT device at magnetic flux $\phii=0.5\phio$, in which case we approximate $\nn=1$ and obtain
\be
\sishot(V) =2\rt^{-1}\int d\epsilon\, \ns(\epsilon)\,\{f(\epsilon-eV)[1-f(\epsilon)]+[1-f(\epsilon-eV)]f(\epsilon)\}. \label{e12}
\ee
Figure~\ref{fig:SI_SQUIPT_0} displays the IV characteristics of such a NIS tunnel junction together with the current noise from Eq.~\ref{e12} calculated at $\te=0.05\kelvin$, assuming the superconducting Al gap $\Delta=200\muev$. For an NIN junction we can further set $\nn=\ns=1$, resulting in
\be
\sishot(V) =2\rt^{-1}\int d\epsilon\,\{f(\epsilon-eV)[1-f(\epsilon)]+[1-f(\epsilon-eV)]f(\epsilon)\}. \label{e12b}
\ee
This can be directly integrated to yield $\sishot=(2eV/\rt)\coth(eV/2\kb\te)$. Two basic cases are then immediately obtained from this expression, namely (i) for $e|V|\ll\kb\te$ the equilibrium thermal noise $4\kb\te/\rt$, and (ii) full shot noise $2e|I|$ in the limit $e|V|\gg\kb\te$.

\clearpage

\subsection*{Temperature dependence of the total current noise}

\section*{Acknowledgements }

We acknowledge Micronova Nanofabrication Centre of Aalto University for providing the processing facilities. We thank  O.-P. Saira and M. Meschke for helpful discussions. The work has been supported by the Academy of Finland Center of Excellence program (project number 284594). J. T. P. acknowledges support from Academy of Finland (Contract No. 275167).

\section*{Author contributions statement}

R. N. J. fabricated the device. R. N. J. and J. T. P. performed the experiments, analyzed the data, and wrote the manuscript. D. S. G. and J. P. P. provided theory support. J. P. P. discussed at all stages of the measurement with R. N. J. and J. T. P.. All authors discussed the results and their implications, and contributed to editing the manuscript.

\section*{Additional information}

The authors declare no competing financial interests.

\end{document}